# Measurement of $\gamma_T$ with the $\gamma_T$ Quads On and Off

Xi Yang, James MacLachlan, and Charles M. Ankenbrandt
*Fermi National Accelerator Laboratory*
Box 500, Batavia IL 60510

## Abstract

An experimental procedure for measuring $\gamma_T$ has been developed and tested in two different measurements, with the $\gamma_T$ quads on and off. The results were compared to MAD calculations. The discrepancy between the measured $\gamma_T$ and the calculated $\gamma_T$ is less than 5%.

## Introduction

There are twelve pulsed $\gamma_T$ quads for faster transition crossing in the Booster, which were designed to change $\gamma_T$ one unit within 100 μs and, according to MAD calculation, they modify the tune less than 0.3 %.[1] It is important to find an experimental procedure to measure the amount of $\gamma_T$ changed by the $\gamma_T$ quads when they are fired. The $\gamma_T$ is well known from the optical design to be 5.447 when the $\gamma_T$ quads are off, and it has been used for checking the precision of the $\gamma_T$ measurement. The discrepancy between the measured $\gamma_T$ and the design parameter is about 2.5%. Also, the discrepancy between the measured $\gamma_T$ change when the $\gamma_T$ quads are pulsed and the calculation using MAD for the same condition is about 15%.

## Method

The direct experimental data for the $\gamma_T$ measurement are the closed orbit from the beam position monitors (BPM) and the rf frequency of the accelerating field from the universal time interval counter. Eq. 1 is used for calculating $\gamma_T$[2]

$$\frac{\Delta C}{C}(t) = \frac{1}{\gamma_T^2} \times \frac{\Delta p}{p}(t) \qquad (1)$$

and eq. 2 is used for calculating $\Delta p/p$:

$$\frac{\Delta p}{p}(t) = \left[ \frac{\Delta C}{C}(t) + \frac{\Delta f_{rf}}{f_{rf}}(t) \right] \cdot [\gamma(t)]^2 \qquad (2)$$

Substituting eq. 2 into eq. 1, we get

$$\gamma_T = \left(\gamma(t) \times \sqrt{\frac{\Delta C}{C}(t) + \frac{\Delta f_{rf}}{f_{rf}}(t)}\right) \bigg/ \sqrt{\frac{\Delta C}{C}(t)} = \gamma(t) \times \sqrt{1 + \frac{\frac{\Delta f_{rf}}{f_{rf}}(t)}{\frac{\Delta C}{C}(t)}} \qquad (3)$$

where $f_{rf}$ is the measured frequency using the HP5370 B universal time interval counter in the control room, $C$ is the circumference of the Booster ring, and $p$ is the momentum of a proton. $\gamma$ is the Lorentz's relativistic factor; $\gamma_T$ is $\gamma$ at which transition occurs. $\Delta f_{rf}$, $\Delta C$, and $\Delta p$ are the changes of the rf frequency, circumference, and proton momentum caused by the change of the radial orbit offset (ROF). The rf frequency ($f_{rf}$) and $\gamma$ for a Booster cycle are calculated from the magnet ramp, and are shown in Fig. 1. According to Fig. 1, the $\gamma$ of a proton at time $t$ can be found from the rf frequency measured at $t$. $\Delta C$ is calculated from the BPM data:

$$\Delta C = \Delta C_1 + \Delta C_2 \qquad (4)$$

where

$$\Delta C_1 = 2 \times \pi \times \left(\frac{\sum_{i=1}^{m} x_i}{48}\right)$$

and

$$\Delta C_2 = \left(\sqrt{(s_{48,1} \times k)^2 + (x_{48} - x_1)^2}\right) - (s_{48,1} \times k) + \sum_{i=1}^{48}\left(\left(\sqrt{(s_{i,i+1} \times k)^2 + (x_i - x_{i+1})^2}\right) - (s_{i,i+1} \times k)\right)$$

Here, $\Delta C_1$ is the circumference change coming from the change of the average beam radius around the ring, the $x_i$ are the horizontal BPM data, and $m=48$ is the total number of BPM's. $\Delta C_2$ is the circumference change coming from the transverse displacement of the beam and the $s_{i,i+1}$ are the longitudinal separations (LS) between the $i^{th}$ and $(i+1)^{th}$ BPM's. $S_{48,1}$ is the LS between the last ($48^{th}$) BPM and first ($1^{st}$) BPM around the ring. $k$ is a scaling factor, which is used to correct the change of LS between two BPM due to the average radius change:

$$k = \frac{\left(R_0 + \left(\frac{\sum_{i=1}^{m} x_i}{48}\right)\right)}{R_0} \qquad (5)$$

where $R_0$=75.472 m is the radius of Booster.

## Results

All the measurements were done at the extracted beam intensity of $0.315 \times 10^{12}$ protons. A programmed radial offset (ROF) was used to move the beam during the time when the rf frequency and the closed orbit were simultaneously measured. The measurement was repeated at eight different ROF values, -2, -1, 0, 1, 2, 3, 4, 5 for the purpose of extracting the slopes of the rf frequency and the change of the circumference with ROF more precisely.

The first measurement was taken at 18.12 ms in the Booster cycle when the $\gamma_T$ quads were off. Here, the beam was injected at 2.0 ms. The rf frequency *vs*. ROF is shown in Fig. 2(a), and a slope of -0.0002 is obtained from the linear curve fit. All the rf frequencies were measured by taking a 40-point average in order to minimize the error from the cycle-to-cycle fluctuation. The closed orbits are shown in Fig. 2(b). The change of the circumference *vs*. ROF is shown in Fig. 2(c), and a slope of -8.1476 is obtained from the linear fit. Eq. 3 is used to find $\gamma_T$=5.5858; the discrepancy with the design parameter $\gamma_T$=5.447 is about 2.55%.

The $\gamma_T$ measurement was repeated at 19.02 ms in a Booster cycle when the $\gamma_T$ quads were fired at 18.92 ms and reached their peak current of 760 A at the 2 kV setting.[3] The rf frequency *vs*. ROF is shown in Fig. 3(a), and a slope of 0.00005 is obtained from the linear fit. The closed orbits are shown in Fig. 3(b). The change of the circumference *vs*. ROF is shown in Fig. 3(c), and a slope of -10.98 is obtained from the linear fit. Eq. 3 is used to find $\gamma_T$=5.2676; the $\gamma_T$ calculated using MAD is 5.2346. The discrepancy between the measured $\gamma_T$ and the calculated $\gamma_T$ is about 0.63 %.

## Conclusion

The direct measurement of $\gamma_T$ was made with the $\gamma_T$ quads on and off, and the results were compared with the calculation using MAD with the $\gamma_T$ quads on and with the design value for the $\gamma_T$ quads off.  Here, the $\gamma_T$ design value is nearly identical to that calculated by MAD.  Since the discrepancy between the measured $\gamma_T$ and the calculated $\gamma_T$ using MAD is less than 5% for either $\gamma_T$ quads on or off, it is satisfactory to adjust the operational parameters for the $\gamma_T$ quads at the high intensity using MAD calculations.

**Acknowledgement**

Thanks to Dr. Alexandr Drozhdin for providing the Booster lattice file and helping the author (X. Y.) perform the MAD calculations of lattice functions at the particular times and parameters relevant to our experiment.

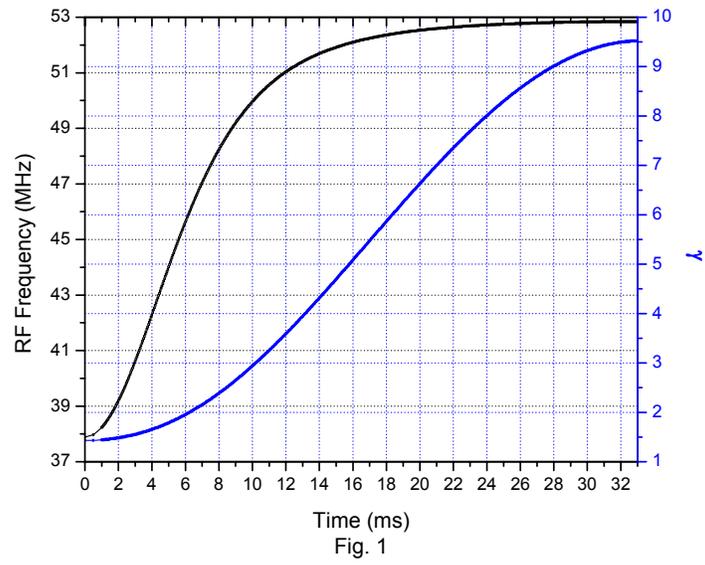

Fig. 1 the calculated rf frequency for a Booster cycle is shown as the black curve, and the calculated Lorentz's relativistic factor ($\gamma$) is shown as the blue curve.

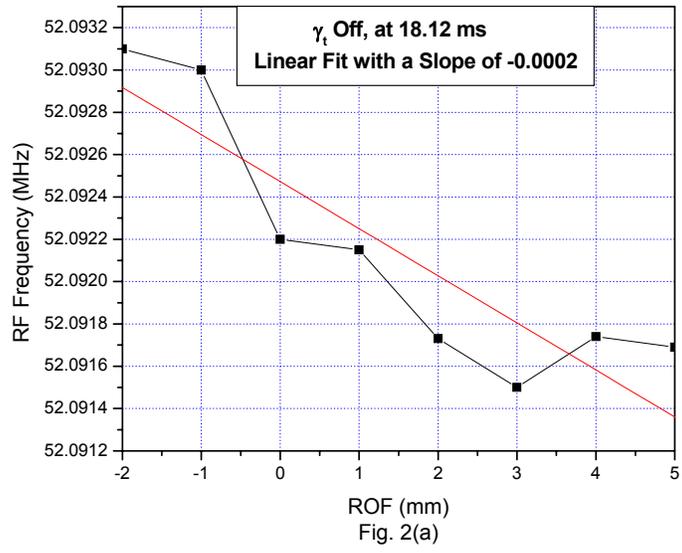

Fig. 2(a)

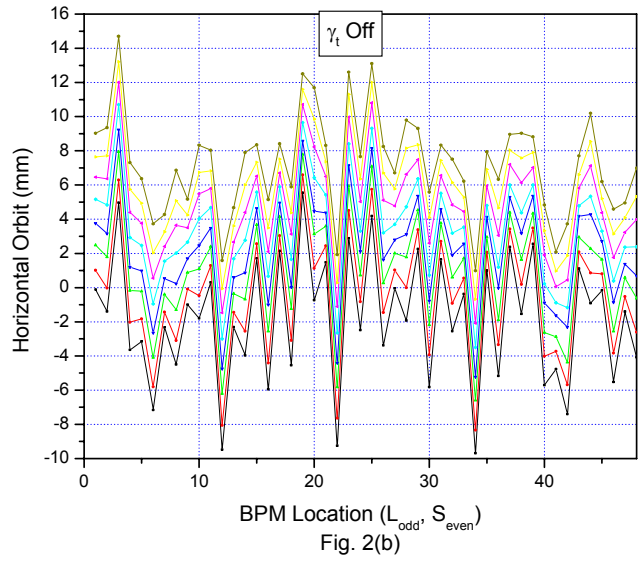

Fig. 2(b)

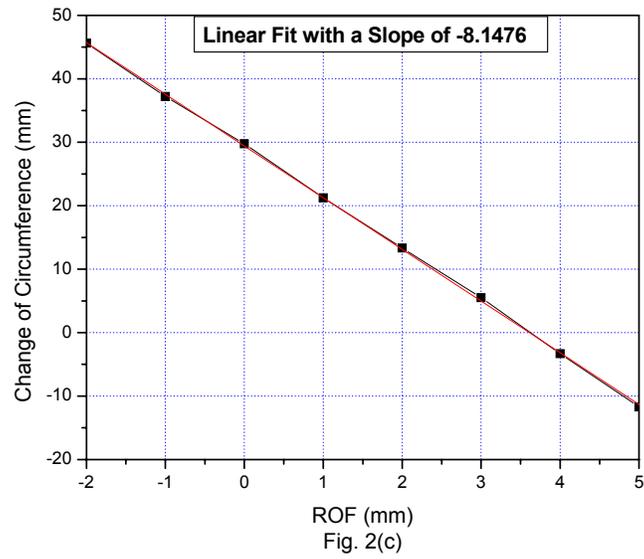

Fig. 2(a) the rf frequency *vs*. ROF measured at 18.12 ms of a Booster cycle for the extracted beam intensity of $0.315 \times 10^{12}$ protons when the $\gamma_T$ quads were off.

Fig. 2(b) the closed orbits measured at eight different ROF values. The black, red, green, blue, cyan, magenta, yellow, dark yellow curves represent the eight different ROF values of 5, 4, 3, 2, 1, 0, -1, -2 respectively.

Fig. 2(c) the change of circumference *vs*. ROF.

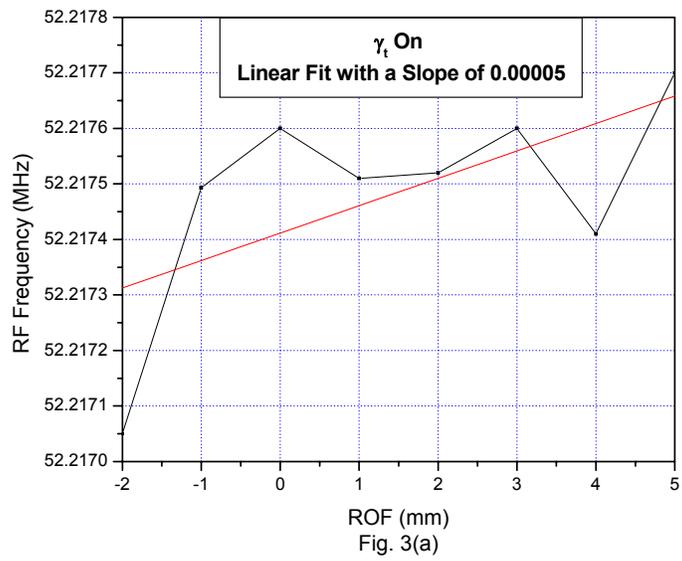
Fig. 3(a)

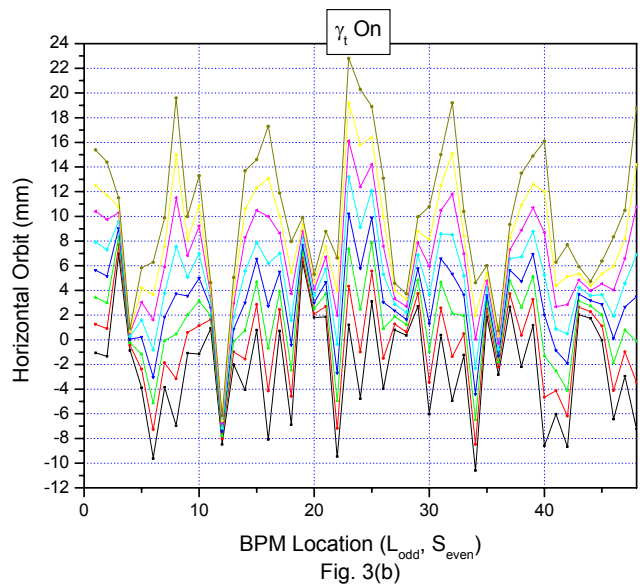
Fig. 3(b)

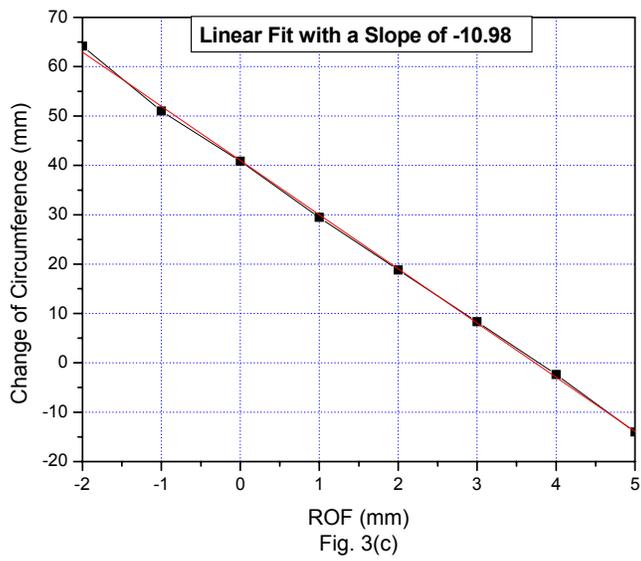

Fig. 3(a) the rf frequency *vs*. ROF measured at 19.02 ms of a Booster cycle for the extracted beam intensity of $0.315 \times 10^{12}$ protons when the $\gamma_T$ quads were turned on at 18.92 ms with a peak current of 760 A.

Fig. 3(b) the closed orbits measured at eight different ROF values. The black, red, green, blue, cyan, magenta, yellow, dark yellow curves represent the eight different ROF values of 5, 4, 3, 2, 1, 0, -1, -2 respectively.

Fig. 3(c) the change of circumference *vs*. ROF.